\documentstyle[12pt]{article}

\begin{document}

\baselineskip = 18pt plus 1pt minus 1pt

\title{About The Bethe--Salpeter--Formalism in the Heavy Mass Limit}

\author{
\makebox[3cm]{}
\\
Achim Wambach\\
\makebox[5cm]{}
\\
Department of Physics\\
Theoretical Physics\\
1 Keble Road, OXFORD OX1 3NP, England\\
\\
\\
OUTP -- 94 01P
\\
}

\vspace{2cm}
\date{1. March 1994}

\maketitle

\begin{abstract}
\noindent
We investigate the Bethe--Salpeter description of mesons in
the limit where one of the constituing quarks is infinitely heavy.
To recover the non--relativistic quark model out of  the
Bethe--Salpeter formalism it is usual to assume that
the potential
is instantaneous and the Fock space is reduced to $Q\bar{q}$ states.
We show that in the Feynman gauge in perturbation theory and for a
heavy  quark being on--mass shell the instantaneity
of the potential is valid in the rest--frame of the meson.
Furthermore we argue that to first order in the velocity exchange during a
heavy meson transition, the error induced by the reduction of the
Fock--space is small.  This in turn justifies quark model calculations, even
when one quark is light.

\end{abstract}

\pagestyle{empty}

\newpage
\pagenumbering{arabic}
\pagestyle{plain}

\section{INTRODUCTION}

The last few years have seen a dramatic progress in the understanding of
heavy quark physics, mainly in the context of a new effective theory.
Originating in the work  by Isgur and Wise
\cite{isgur1}, many people have contributed to complete the
description of Heavy Quark Effective Theory (HQET)
\cite{neubert1}.

However, although the effective theory provides new symmetries and leads to
a deeper understanding of heavy quarks physics, it is not in its range
to describe the physics of the light degrees of freedom in a heavy hadron.

A good example for this shortcoming is the so-called ``Isgur--Wise
function'', which is extensively discussed in the literature
\cite{blok}--\cite{kalinovsky}. The Isgur--Wise function $\xi$ describes
the form factor
for heavy meson decays in the infinite mass limit:
\begin{equation}
\frac{<D(v')|\Gamma_\mu|(B(v)>}{\sqrt{M_D M_B}} = \xi(vv') (v_\mu+v'_\mu)
\end{equation}

Using flavour symmetry, i.e. the light degrees of freedom cannot distinguish
between a heavy b quark and a heavy c quark, one can
relate the expression in (1) to a conserved current expression for
B--mesons only. The consequence is the normalization of the
Isgur--Wise function to one at $v=v'$, the zero recoil point.
However, in order to evaluate experimental data and eventually to extract the
$V_{bc}$ CKM--matrix element, it is necessary to understand more about
$\xi(vv')$, in particular in the region where $v\not=v'$. Investigating
this problem for a small velocity change, formally the first order
term in an expansion of $\xi$ in $(y-1)$ where $y=vv'$, one has to come
back to other methods, HQET alone cannot provide conclusive clues
about the Isgur--Wise function in this region. This effect is only
attributed to the light degrees of freedom.

Several groups attempted to
extract the so called charge--radius $\rho$, which is defined by
$\xi(vv') = 1 - \rho^2(y-1) +O( (y-1)^2 )$.
Methods which were employed are ranging from QCD sum rules
\cite{blok,neubert2,radyushkin,ball}  ,
optical sum rules \cite{voloshin},  model calculations
\cite{close,neubert3,isgur2} to lattices \cite{ukqcd} and the Bethe--Salpeter
formalism \cite{dai1,kugo,kalinovsky}. The results are
ranging from 0.8 \cite{blok} to $\infty$ \cite{radyushkin}, while the
experimental data seems to support a value for $\rho$ of about $1.2 \pm 0.3$
\cite{argus}.
The variation in the theoretical data and the
unique feature of this problem, namely that we are dealing with a single
constituent quark  makes it worthwhile to further investigate
this problem.

In this paper we discuss the description of heavy mesons in the
Bethe--Salpeter formalism \cite{bethe}. This formalism has the advantage
that it provides a relativistic covariant description of moving bound states.
In particular we are interested in the role the light antiquark is playing in
the meson and during the decay.

Quark models do provide us with the physical interpretation of a
spectator--antiquark. The spectator is not affected by the heavy quark
transition during a heavy meson decay. The origin of the form--factor is solely
the wavefunction overlap of the light antiquark. However, this picture  itself
cannot be complete -- effects of retarded potentials are ignored and the
Fock--space in quark models is limited to $Q\bar{q}$ states only.

As we will argue in this paper, in the infinite mass limit of one of the
quarks, which has to be on--mass--shell, the
form of the potential simplifies considerably in the rest
frame of the meson. This enables us to justify an instantaneous
potential, which is at the origin of the reduction of the
Bethe--Salpeter equation to the quark-model description.

In this framework we show also that for heavy meson--transition the
reduction of the Bethe--Salpeter formalism to the quark--model
description is correct, if no  velocity is exchanged, i.e. the zero--recoil
point. For a small
velocity--exchange we discuss the error which arises when the light antiquark
is forced to have positive energy only  and illustrate this effect
in a toy model.  We argue that for a limited size of the meson wavefunction
this effect is surpressed.

This paper is constructed as follows. In section 2 we develop the
necessary formalism in the heavy mass limit, transitions of heavy
mesons in the zero--recoil limit are considered.  In the 3rd section
we investigate moving
mesons and discuss their additional features. We finish with
conclusions in section 4.

\section{THE BETHE--SALPETER FORMALISM}

To develop the Bethe--Salpeter (B--S) formalism in the heavy mass
limit we are following mainly the derivation  given in  \cite{dai1,dai2}.
However, as our
aim is to investigate the role of the light antiquark, we do take its
four--momentum as the free momentum and  not the relative momentum.
The B--S equation for  a meson wave--function is given by:
\begin{equation}
\chi_P(k) = S_1(p) \int \frac{{\rm d}^4q}{(2\pi)^4} G(P, k, q)
\chi_P(q) S_2(-k)
\end{equation}
Here, $p$ is the momentum of the (heavy) quark with mass $M_1$, $P=Mv$
is the overall momentum of the meson with mass $M$ and velocity $v$,
and $k$ is the momentum of the (light) antiquark with mass $m$.

In the heavy mass limit the propagator for the heavy quark reduces to:
\begin{equation}
S_1(p) = \frac{i(\not{\!p}+M_1)}{p^2-M_1^2}  \rightarrow
\frac{1+\not{\!v}}{2} \frac{i}{(M-M_1)-k \cdot v}
\end{equation}
while the propagator for the light antiquark cannot be simplified:
\begin{equation}
S_2(-k) = \frac{i(m-\not{\!k)}}{k^2-m^2} =
\frac{i(m-\not{\!k})}{k_l^2+k_t^2-m^2}
\end{equation}
where we introduced $k_l = k \cdot v $ and $ k_t = k - k_l v$.
The reduction of the heavy quark propagator becomes more comprehensible if
one uses positive and negative energy--states projectors. In the rest
frame of the meson they have the form:
\begin{equation}
\Lambda_{\pm}(\vec{p}) = \frac{E(\vec{p}) \pm H(\vec{p})}{2 E(\vec{p})}
\end{equation}
where $E(\vec{p}) = \sqrt{M_1^2 + \vec{p}^2}$ and  $H(\vec{p}) = \beta
M_1 + \vec{\alpha}\vec{p}$. The propagator in eq.(3) then reduces to:
\begin{equation}
S_1(p) = i\frac{\Lambda_+(\vec{p})\gamma_0}{p_0- E + i\epsilon} +
i\frac{\Lambda_-(\vec{p})\gamma_0}{p_0 + E -i\epsilon} \rightarrow
i\frac{\Lambda_+(\vec{p})\gamma_0}{p_0- E+i\epsilon}
\end{equation}
or in an arbitrary frame:
\begin{equation}
S_1(p) =   \frac{i}{2\omega_1} (\frac{M_1+\omega_1 \not{\!v} +
\not{\!k_{t}}}{ (M-\omega_1)-k_l+i\epsilon} + \frac{-M_1+\omega_1
\not{\!v} + \not{\!k_{t}}}{(M+\omega_1)- k_l-i\epsilon}) \rightarrow
\frac{1+ \not{\!v}}{2} \frac{i}{(M-M_1)-k_l+i\epsilon}
\end{equation}
with $\omega_1 = \sqrt{M_1^2 - p_t^2}$. The heavy mass limit therefore
picks out the quark part of the heavy quark and eliminates the
antiquark part.

It is worthwhile to note that in this limit only the term with
$k_l-i\epsilon$ in the denominator survives. This will be of crucial
importance for the reduction of the Fock-space to $Q\bar{q}$ states
only.

By using the most general form for the Kernel  \cite{dai1}
\begin{equation}
-iG(P,k,q) = 1 \otimes 1 V_1(v,k,q)  +  v_\mu \otimes \gamma^\mu V_2(v,k,q)
\end{equation}
and for the pseudoscalar meson, to which we are going to restrict our
discussion
(extensions to other states have been done elsewhere
\cite{dai2,wambach,hussain})
\begin{equation}
\chi_P(k) = \gamma_5 (1-\not{\!v}) ( \phi_1(k,v) + \not{\!k}_t \phi_2(k,v))
\end{equation}
one arrives at the following defining equations for $\phi_1$ and $\phi_2$:
\begin{equation}
\begin{array}{ccc}
\phi_1(k) = \alpha \int \frac{{\rm d}^4 q}{(2\pi)^4} [\phi_1(q)
(V_1(k,q)-V_2(k,q)) (m+k_l) - \phi_2(q) (V_1(k,q) + V_2(k,q)) q_t \cdot k_t]\\
\\
\phi_2(k) = \alpha \int \frac{{\rm d}^4 q}{(2\pi)^4} [- \phi_1(q)
(V_1(k,q)-V_2(k,q)) + \phi_2(q) (V_1(k,q) + V_2(k,q)) \frac{q_t \cdot
k_t}{k_t^2} (m-k_l)]
\end{array}
\end{equation}
where $\alpha = -i [(M-M_1-k_l+i \epsilon)(k_l-\omega + i\epsilon) (k_l
+ \omega - i\epsilon)]^{-1} $.

It is usual at this stage to assume that the potential is instantaneous
in the rest frame of the meson, i.e. $V_i(k,q) = V_i (|\vec{k} -
\vec{q}|)$ $(i=1,2)$ to solve these equations
\cite{dai1,kugo,kalinovsky,dai2}.
Restricting  to the Feynman gauge for the Kernel in perturbation
theory and assuming that the heavy quark is on--mass--shell one can
derive this instantaneous behaviour directly (following \cite{flamm}).
An instantaneous potential for the meson in its rest frame can also be
derived in the Coulomb gauge for the gluon propagator, independent of
the mass of the constituing quarks. However, in contrast to the
Feynman gauge, the Coulomb gauge fixing is not covariant. It is this
covariance of the Feynman gauge which will enable a consistent
description of moving mesons.
In Feynman gauge the
gluon-propagator and the two vertex--functions with the heavy and the
light quark take the  form (with $\rho = k-q$):
\begin{equation}
-i G(P, \rho) =  \frac{16\pi}{3}
\frac{\alpha_s(-\rho^2)}{(2\pi)^4} \gamma_{\mu} \otimes
\gamma_{\nu} \frac{g^{\mu \nu}}{\rho^2}
\end{equation}
$\rho$ is the four--momentum of the exchanged gluon, and an additional
factor of
$\frac{4}{3}$ is included because summation over colour is assumed.
For the heavy quark being
on--mass-shell, current conservation applies on the heavy quark line,
i.e.:
\begin{equation}
\rho^{\mu}\gamma_{\mu} \otimes \gamma_{\nu} = 0
\end{equation}
Using the following equalities:
\begin{equation}
\frac{\gamma_0 \otimes \gamma_0}{\rho^2} = \frac{\gamma_0
\otimes \gamma_0}
{\vec{\rho}^2} \frac{ \rho_0^2 - \rho^2}{\rho^2} = -
\frac{\gamma_0 \otimes \gamma_0}{\vec{\rho}^2} +
\frac{\rho_0^2\gamma_0 \otimes \gamma_0}{\rho^2 \vec{\rho}^2}
\end{equation}
we derive:
\begin{equation}
-iG(P,\rho) \rightarrow  -\frac{16\pi}{3} \frac{\alpha_s(-\rho^2)}{(2\pi)^4} \{
\frac{\gamma_0 \otimes  \gamma_0}{\vec{\rho}^2} + \frac{1}{\rho^2} [
\vec{\gamma} \otimes \vec{\gamma} -
\frac{(\vec{\rho}\vec{\gamma}) \otimes (\rho_0\gamma_0)}{\vec{\rho}^2} ] \}
\end{equation}

In the rest--frame of the heavy quark, the propagator of the heavy quark
line has the form $\frac{1+\gamma_0}{2}$. Sandwiching the gluon--heavy quark
vertex as derived above between two rest--frame projectors, it
immediately follows that only the first term in (14) survives. If we
neglect the $\rho^2$ dependence of $\alpha_s$
the  gluon propagator is proportional only to $\vec{\rho}$, the
three-momentum; the Kernel is instantaneous.

However, the assumption that the heavy quark is on--mass--shell is not
obvious and  not correct in HQET. With the
four--momentum of the heavy quark as $p=M_1v + \kappa $ the
off--shellness is $p^2 - M_1^2 = 2M_1 v \cdot \kappa + $O(1), which
is of the order $M_1$ and does therefore not diminish in the heavy mass
limit.\footnote{The author is indebted to A. Le Yaouanc for pointing that
out.} A more careful investigation of this effect is necessary but goes
beyond the scope of the present analysis.

 From now on we want to
restrict ourselves to the use of an
instantaneous potential in the rest frame of the meson in order to elicit
the role of the light antiquark.

An instantaneous potential in the rest frame of the meson is
`covariant instantaneous' \cite{sadzijan} in an arbitrary frame, i.e.
$V_i(k, q) = V_i(k_t, q_t)$, $(i=1,2)$. This comes about because the rest
frame--potential in real space transforms like
$V_i(x_\mu)=V_i(\vec{x})\delta(x^0) =V_i(x_\mu)\delta(x.v)$ $(i=1,2)$.
Fourier transforming this expression the constraint $k_l-q_l=0$ follows.

With these remarks in advance, eqs.(10) transform to:
\begin{equation}
\begin{array}{c}
\hat{\phi}_1 (k_t) = - (m+\omega)\hat{\phi}_2(k_t)\\
\\
\hat{\phi}_2 (k_t) =  \beta \int \frac{{\rm
d}^3q_t}{(2\pi)^3}\hat{\phi}_2(q_t)[
(m+\omega_q)(V_1(k_t,q_t)-V_2(k_t,q_t)) + \frac{q_t \cdot k_t}{k_t^2}
(V_1(k_t,q_t)+V_2(k_t,q_t))(m-\omega)]
\end{array}
\end{equation}
where $\beta=-[2\omega(M-M_1-\omega)]^{-1}$,
$\omega_q=\sqrt{m^2-q_t^2}$, $\omega=\sqrt{m^2-k_t^2}$ and
$\hat{\phi}_i(q_t)=\int \frac{{\rm d}q_l}{2\pi}\phi(q)$.

An interesting result appears if the meson--wavefunction is reformulated
slightly, namely :
\begin{equation}
\chi_P(k) = \gamma_5 (1-\not{\!v}) ( \Phi_1(k,v) + \not{\!k} \Phi_2(k,v))
\end{equation}
with $\Phi_1(k) = \phi_1(k) + k_l \phi_2(k)$ and $\Phi_2(k) = \phi_2(k)$.
The defining equation for $\hat{\Phi}_1$ in the covariant instantaneous
potential approximation is then:
\begin{equation}
\hat{\Phi}_1(k) = -m \hat{\Phi}_2(k)
\end{equation}
and therefore the wave--function transforms to:
\begin{equation}
\chi_P(k) = \gamma_5 (1-\not{\!v}) ( -m + \not{\!k}) \hat{\Phi}_2(k,v)
\end{equation}
This is exactly the form which was derived in \cite{close,wambach} by
starting from a non--relativistic quark--model, and where the
Wigner--rotation effects of the light antiquark were included consistently.\\

Having set up the formalism of the B--S wave--function, we now turn to
the decay of one heavy meson into another, first in the zero--recoil
limit. To simplify, we only discuss mesons of the same type, which has
the effect that  $E \equiv M-M_1-m $ is always the same quantity.
Because of the scaling properties of the heavy mass limit, namely
$\langle M'(v')|\Gamma_\mu|M(v) \rangle = \sqrt{MM'} \xi(v\cdot v')
(v+v')_\mu$ with $\xi$  independent of the heavy mass, effects
attributed to a different $E'=M'-M_1'-m$ should scale out. Therefore it
is sufficient to restrct attention to mesons with the same $E$.

For $v=v'$, the decay element is \cite{leyaouanc}:
\begin{equation}
\begin{array}{lc}
\lefteqn{ \langle M'(v)|\Gamma_\mu|M(v) \rangle  =  i \int \frac{{\rm
d}^4k}{(2\pi)^4} [\bar{\chi}'(\gamma_\mu \otimes (\not{\!k}+m)) \chi] } \\
\\
=  & \int \frac{{\rm d}^4k}{(2\pi)^4}  \frac{{\rm d}^4 q}{(2\pi)^4}
\frac{{\rm d}^4q'}{(2\pi)^4}
\frac{-i}{(E+m-k_l+i\epsilon)^2(k_l-\omega+i\epsilon)(k_l+\omega-i\epsilon)}
\times\\
\\
&  {\rm Tr}[(V_1'+\not{\!v}V_2')(\phi_1'+\not{\!q}_t' \phi_2')
(1-\not{\!v}) \gamma_\mu (1-\not{\!v}) (\phi_1+\not{\!q}_t \phi_2)
(V_1+\not{\!v}V_2) (m-\not{\!k})]
\end{array}
\end{equation}
where the unprimed (primed) quantities are functions of $q$($q'$).
In this equation $m$ describes the renormalized mass of the light
quark, as e.g. discussed in \cite{flamm}. We assume that the mass of
the light quark does not change during the transition.

Using the instantaneous approximation we may integrate over d$k_l$,
d$q_l$ and d$q_l'$.
Of particular interest is the integration over $k_l$, the covariant
energy of the light antiquark. The pole structure of the expression
reveals that integration over $k_l$ picks up only one pole, namely
$k_l=\omega$, if we close the contour at the lower half of the
complex plane. In the projector picture as developed above,
\begin{equation}
(m-\not{\!k}) =\frac{1}{2\omega}[(m-\omega\not{\!v}-\not{\!k}_t)
(k_l+\omega) + (-m -\omega\not{\!v}+\not{\!k}_t)(k_l-\omega)]
\end{equation}
only the first term
survives. The fact that the heavy quark is set to be a positive energy
state forces the light antiquark to have positive energy as well, or,
equivalently, the Fock--space is reduced to $Q\bar{q}$ states only.

Evaluating the trace in eq.(19) the following equation turns out to be true:
\begin{equation}
\begin{array}{ccc}
 \langle M'(v)|\Gamma_\mu|M(v) \rangle & = & i \int \frac{{\rm
d}^4k}{(2\pi)^4} [\bar{\chi}'(\gamma_\mu \otimes (\not{\!k}+m)) \chi] \\
\\
& = &  - \int \frac{{\rm d}^3k_t}{(2\pi)^3}(\frac{\omega}{m})
[ \hat{\bar{\chi}}'(\gamma_\mu \otimes 1) \hat{\chi}]  \\
\end{array}
\end{equation}
where
$\hat{\chi}=\gamma_5(1-\not{\!\!v})(-(m+\omega)+\not{\!\!k}_t)\hat{\phi}_2$.
This consolidates the equivalence of the B--S formalism in the
covariant instantaneous approximation and the quark model formalism
developed in \cite{close,wambach}. There the integration is done over
the three--momentum only while the energy of the light antiquark is kept
positive, $k_l = + \sqrt{m^2-k_t^2}$.

The discussion of the behaviour of the light quark given above can
also be done graphically. In the rest frame of the heavy meson, the
potential is instantaneous, which means that as
plotted in figure 1.1. the interaction--line has to be vertical.
Figure 1.1. shows the heavy quark transition while the light antiquark
keeps its positive energy (no Z--diagrams). Diagrams like figure 1.2
are forbidden, because here the potential is not instantaneous.

\input epsf
\vspace*{-11cm}
\epsfbox{diagram1.ps}
\vspace*{-2.6cm}
    {\it{ {\rm[1]} Figure 1.1  shows the heavy quark interaction with
an external current, while the interaction with the light antiquark
is done via an instantaneous potential. Figure 1.2 is not allowed,
because here the potential is not instantaneous.}}

\vspace{0.8cm}

In figure 2.1 it appears to be the case that in this formalism  a transition
of a negative energy state to a negative energy state on the light
quark line is allowed. However, plotting this diagram in Feynman form
(fig. 2.2) reveals that this is a correction to the meson--current, not to the
wave--function, which we do not want to discuss here.

\input epsf
\newpage
\vspace*{-15cm}
\epsfbox{diagram2.ps}
\vspace*{-2.4cm}
{\it{ {\rm [2]} In Figure 2.1. a possible candidate for a
negative $\rightarrow$ negative energy state transition is shown.
However diagram
2.2 in Feynman form reveals that this is a correction to the meson--current
and not to the wave--function.}}
 \vspace{0.8cm}

It is important to note that this reduction of the Fock--space, namely that the
light antiquark is fixed to positive energy states, is only
valid if the interaction takes place on the heavy quark line. For
interactions involving the light quark (e.g. electromagnetic
interactions) transitions between negative--energy and
positive--energy states are possible, as displayed in figure 3.1,
3.2., although negative $\rightarrow$ negative energy states
transitions are still
not possible (figure 3.3).

\input epsf
\vspace*{-14.5cm}
\epsfbox{diagram3.ps}
\vspace*{-2.1cm}
   {\it{ {\rm [3]}Diagrams 3.1 and 3.2 show the transition from a
negative--energy  to a positive--energy light antiquark state and
vice versa, which  happens if
the interaction takes place with the light antiquark. In diagram 3.3.
it is indicated that negative $\rightarrow$ negative energy transitions
are not allowed.}}
 \vspace{0.8cm}

The reduction of the wave--function from a four--dimensional
 to a three dimensional integral is very helpful by determining
kinematical quantities like the spectroscopy. The extension to
dynamical quantities like form factors is not trivial. We have
shown in this section that in the zero recoil limit this extension is valid. We
now
turn to meson--transitions where the velocity exchange is non--zero,
but small.

\section{MOVING MESONS}

Before setting up the formalism we shall discuss the possible outcome
diagrammatically. If the meson is at rest before the transition  and
moves  with velocity $v'$ in the direction of the
z--axis (for simplicity) after the transition, possible transition diagrams
have the form
of figure 4.1. Note that the potential is not instantaneous after the
transition.

A type of diagram, which might destroy the wanted feature that the
light antiquark keeps its positive energy, is displayed in figure
4.2.

\input epsf
\vspace*{-12.7cm}
\epsfbox{diagram4.ps}
\vspace*{-2.1cm}
  {\it{ {\rm [4]} This diagram shows the transition of a meson in its
rest--frame to a moving meson, where the potential is
retarded. In 4.2 the light antiquark does pick up negative--energy
states. [$X_i=(T_i, \vec{X}_i)$, $x_i=(t_i, \vec{x}_i)$]}}
  \vspace{1cm}

\newpage
However, using the notation of figure 4.2 the following constraints
for this transition can be derived:
\begin{equation}
\begin{array}{rcl}
T_2 & > & T_1 = t_1\\
t_2 & < & t_1 \\
(T_2 - t_2)v_0 & = & (Z_2 - z_2)v\\
(T_2 - T_0)v & = & (Z_2 - Z_0)\\
\end{array}
\end{equation}
$Z_0 = Z_1$ is the
interaction--place $z$--component, $T_0 > T_1$ is the interaction--time.
For the third equation it is used that in the rest frame of the meson after the
transition the potential has to be instantaneous. Therefore the retardation of
the potential in the moving frame is limited.
The last equation displays the fact that the heavy quark is assumed to
be moving classically
after the transition. This than yields
for the distance of the light quark from the heavy quark and therefore
from the center of the meson:
\begin{equation}
|z_2 - Z_2| = |t_2 - T_2|\frac{v_0}{v} > |T_1 - T_2|\frac{v_0}{v} >
|Z_2-Z_1|\frac{v_0}{v^2}
\end{equation}
The appearance of $v$ and $v^2$  in the denominator indicates that this
has to take place far away from the center of the meson. If the meson
wave--function is localized in space as it is in QCD due to its confining
properties this effect should be surpressed. As we shall
discuss later on, a constant vertex function, for example, which is
equivalent to a
point-like vertex, effectively forbids this effect.

The diagrammatic result is therefore that corrections to the usual
three--dimensional reduction of the integration in the B--S
wavefunction due to negative--energy states should be small if the
velocity--exchange is small. We now want to show this formally by
direct inspection of the B--S formalism for meson transitions.

The decay element in its most general form is:
\begin{equation}
\begin{array}{lc}
\lefteqn{\langle M'(v')|\Gamma_\mu|M(v) \rangle  =  i \int \frac{{\rm
d}^4k}{(2\pi)^4} [\bar{\chi}'(\gamma_\mu \otimes (\not{\!k}+m)) \chi]} \\
\\
 = & \int \frac{{\rm d}^4k}{(2\pi)^4}  \frac{{\rm d}^4 q}{(2\pi)^4}
\frac{{\rm d}^4q'}{(2\pi)^4}
\frac{-i}{(E+m-k_l+i\epsilon)^2(k_l-\omega+i\epsilon)(k_l+\omega-i\epsilon)}
\times\\
\\
&  {\rm Tr}[(V_1'+\not{\!v'}V_2')(\phi_1'+\not{\!q}_t' \phi_2')
(1-\not{\!v'}) \gamma_\mu (1-\not{\!v}) (\phi_1+\not{\!q}_t \phi_2)
(V_1+\not{\!v}V_2) (m-\not{\!k})]
\end{array}
\end{equation}
As discussed above, we used $E'+m=E+m$, and note that
$(k_l'-\omega')(k_l'+\omega') = (k_l-\omega)(k_l+\omega)$. In this case
$q_t'=q'-(q'\cdot v')v'$ while $q_t=q-(q\cdot v)v$.

In the heavy mass limit this expression has to be proportional to
$(v+v')_\mu$ \cite{neubert2}. We therefore multiply it by $(v+v')^\mu$
which contracts with $\gamma_\mu$
to -2  inside the trace. Using the covariant instantaneous
approximation, eq.(24) can be reduced further:
\begin{equation}
\begin{array}{c}
(v+v')^\mu \langle M'(v')|\Gamma_\mu|M(v) \rangle =
\int \frac{{\rm d}^4k}{(2\pi)^4}
\frac{2i}{(E+m-k_l+i\epsilon)^2(k_l-\omega+i\epsilon)(k_l+\omega-i\epsilon)}
\times\\
\\
 {\rm Tr}[(-\Gamma_1'+\not{\!k}_t'\Gamma_2')
(1-\not{\!v}')(1-\not{\!v}) (-\Gamma_1+\not{\!k}_t \Gamma_2)
(m-\not{\!k})]
\end{array}
\end{equation}
where
\begin{equation}
\begin{array}{ccc}
\Gamma_1^{(')}(|k_t^{(')}|) & = & \int \frac{{\rm d}^3 q_t^{(')}}{(2\pi)^3}
(m+\omega_{q^{(')}})(V_1-V_2)(|k_t^{(')}-q_t^{(')}|)
\hat{\phi}_2(|q_t^{(')}|)\\
\\
\Gamma_2^{(')}(|k_t^{(')}|) & = & \int \frac{{\rm d}^3 q_t^{(')}}{(2\pi)^3}
\frac{q_t^{(')}\cdot k_t^{(')}}{k_t^{(')2}}
(V_1+V_2)(|k_t^{(')}-q_t^{(')}|) \hat{\phi}_2(|q_t^{(')}|)
\end{array}
\end{equation}

The evaluation of this expression with different Ans{\"a}tze for
$V_1$ and $V_2$ has been done elsewhere \cite{dai1,kugo,kalinovsky,dai2}.
In the present analysis we want to discuss the implication this
equation has for the light antiquark and  therefore do not  calculate concrete
numbers.

As a first set of possible vertex structures
suppose the vertices $\Gamma_i$ $(i=1,2)$ are just constants.
Plugging  these
vertices into equation (25) one encounters no further pole in $k_l$, so
that the antiquark keeps its positive energy during the transition. The reason
for this behaviour is evident: The Fourier
transform of these vertices is $\Gamma_i(\vec{r}) = \delta(\vec{r})$.
The vertices
are therefore not only instantaneuos (in the rest frame), but also pointlike.
The physical
picture of the distance of the light antiquark from
the meson for possible negative--energy contributions, as given above,
is valid. The
spatial restriction of the vertices  effectively forbids this contribution.

Similar
results apply if the vertices are proportional to $k_t^n$ with $n>0$.
In real space these  vertices are proportional to $\frac{{\rm d}^n}{{\rm d}r^n}
\delta(\vec{r})$, the vertices are again pointlike and the same
arguments apply.

The situation is different if one encounters possible new poles. In
particular rest--frame vertex--functions like $\Gamma_i(\vec{k}) = \sum b_n
((\vec{k})^2+\mu^2)^n$, where $n<0$ can create additional effects.
To illustrate these effects and in particular their mathematical origin
we employ a toy model for the vertex functions which
allows an analytical  calculation of the decay element  and in the same
time keeps contact with the necessary physical features.

This model is inspired by the Feynman gauge as given above. In that
gauge it follows that $V_1=0$ and $V_2 \propto \frac{1}{\vec{k}^2}$ in
the rest frame of the decaying meson (in which we work from now on) and
therefore
\begin{equation}
\begin{array}{ccc}
\Gamma_1(\vec{k}) & \propto & - \int \frac{{\rm d}^3q}{(2\pi)^3}
(m+\omega_q)\frac{1}{(\vec{k}-\vec{q})^2} \hat{\phi}_2(\vec{q})\\
\\
\Gamma_2(\vec{k}) & \propto & \int \frac{{\rm d}^3q}{(2\pi)^3}
\frac{\vec{k}\cdot \vec{q}}{\vec{k}^2} \frac{1}{(\vec{k}-\vec{q})^2}
\hat{\phi}_2(\vec{q})
\end{array}
\end{equation}

We do not attempt to solve these equations. However, inspired by the
form of the
defining equation for $\Gamma_i$, $(i=1,2)$ we define the vertices to
be:
\begin{equation}
\begin{array}{ccc}
\Gamma_1(\vec{k}) & = & -\sqrt{Mm}\frac{2m}{\vec{k}^2+\mu^2}\\
\\
\Gamma_2(\vec{k}) & = & \sqrt{Mm} \frac{1}{\vec{k}^2+\mu^2}
\end{array}
\end{equation}
The mass $\mu$
is included to avoid infrared
singularities in the calculation.

In the pole structure given in eq.(25) there is in the lower half of
the complex plane only one pole for $k_l$, namely $k_l = \omega
-i\epsilon$. The $\Gamma_i$ do only depend on  $ k_t$ so that they
cannot contribute further pole terms . However the vertices in the
boosted frame, $\Gamma _i'$ depend on $k_t^{2'}$ which is equal to
$(k-(k\cdot v') v')^2 = k_t^2-(k_t\cdot v')^2 + (1- y^2 )k_l^2 -2yk_l(k_t\cdot
v')$ where $y=v\cdot v'$ and they
might contribute further poles in $k_l$ (or $k_0$ in the rest frame).

In the rest frame of the decaying meson and with the meson moving
after the decay in the direction of the z--axis, this additional pole
is given for
\begin{displaymath}
k_0 = \frac{1}{v} (k_zv_0+i\sqrt{k_x^2+k_y^2+\mu^2}) \equiv \frac{1}{v}B
\end{displaymath}

Counting the appearance of $k_0$ in the numerator and denominator in
eq.(25) it turns out that only the terms with $k_0$ in the numerator
do contribute to the pole in order $v^2=2(y-1)+$O($v^4$). These are
$\Gamma_1'\Gamma_1(-k_l)+\Gamma_2'\Gamma_2(k_t'\cdot k_t)k_l$. It is
exactly here where the new contributions of that additional pole
appear. Here the projector of the negative--energy state of the light
antiquark is not zero.
In particular, both terms $\frac{1}{2\omega}[(m-\omega\not{\!v}-\not{\!k}_t)
k_l$ and $\frac{1}{2\omega} (-m -\omega\not{\!v}+\not{\!k}_t)k_l$ do
count (cf. eq.(20)).

Other terms , like these given below, have $k_0$ accompanied by an additional
factor $v$, so that they add only to  higher order contributions.
These terms are:
\begin{displaymath}
\begin{array}{ccc}
(k_t\cdot v')k_l & \propto & k_zv'k_0\\
\\
(k_t'\cdot v) & \propto & k_0(1-y^2)\\
\\
(k_t'\cdot k_t) & \propto & k_zv'k_0y
\end{array}
\end{displaymath}

Evaluating eq.(25) at the additional pole then yields:
\begin{equation}
 4 v^2 M(1+y) \int\frac{{\rm d}^3 k}{(2\pi)^3}
\frac{4m^2+\vec{k}^2-Bk_z}{B^3(\vec{k}^2+\mu^2)(\sqrt{k_x^2+k_y^2+\mu^2})} \\
\\
=  2 M (1+y) (y-1) \frac{1}{24\pi} \frac{m}{\mu} (5+4\frac{m}{\mu}^2)
\end{equation}

To determine the relative value of this  additional pole in that toy model one
first has to determine $(E+m)$, $m$ and $\mu$. $(E+m)$ was estimated in
\cite{dai1,dai2} to be between 105MeV and 310MeV (with a strong dependency on
the $B^*$ mass). Following \cite{isgur2} we use $m=330$MeV.  By the constraint
$\xi(1)=1$, $\mu$ is then fixed and lies between 185MeV and 535MeV, which is in
the range of values one would expect if $\mu$ was interpreted as the mass in a
Yukawa potential in a non--relativistic model. The charge radius $\rho$ of the
form factor turns out to be between $ 0.68(-0.05) $ and $ 1.29(-0.18)$, where
the contribution of the additional pole is given in the parentheses.

However, the numeric value  has to be considered with caution. A change in the
definition of $\Gamma_i$ , ($i=1,2$), by a factor of 2 ($\frac{1}{2}$) changes
the average value of $\rho$ to 0.87 (1.36). The model is employed solely to
illustrate the origin of the additional pole and with that the cause of
negative--energy contributions of the light antiquark.

\section{CONCLUSIONS}

We have argued that in the case of an instantaneous potential in the
Bethe--Salpeter equation the error induced by the reduction of the meson
wavefunction to a three dimensional integral is small if the velocity exchange
during the heavy meson transition is small.

This is induced by the fact that negative--energy contributions of the light
antiquark to the decay of the heavy meson are surpressed by the finite volume
of the meson--wavefunction. With constant vertex functions and vertex functions
proportional to $\vec{k}^n$ with $n >0$ negative--energy contributions did not
appear. To illustrate the effect in  other forms of the vertex function we
performed a calculation in  a Yukawa--type toy model.
This model showed a consistency in so far as that for a given value of the
binding energy, the Yukawa  mass and the charge radius are all in a range of
sensible values. However, the intention of the use of this model was to show
were exactly this additional pole--term arises. The numeric results have to be
regarded with caution.

It was also shown that in the case of a heavy meson being on--mass--shell, the
reduction to an instantaneous potential in the rest frame of the meson is valid
in perturbation theory. Further work has to be done to investigate to what
respect the off--shellness of the heavy quark in HQET might change this result.
\vspace{1cm}

It is a pleasure for me to thank Olivier
P{\`e}ne, Alain Le Yaouanc and Jean--Claude Raynal for many fruitful and
helpful discussions. Furthermore I would
like to thank the Theory Group in Orsay for their hospitality I enjoyed during
the preparation of parts of the present work.
Support from the German Academic Exchange Service and the German Scholarship
Foundation is gratefully acknowlegded.

\end{document}